\documentclass{article}
\usepackage{graphicx}
\usepackage{times}
\usepackage[left=2.5cm,top=2.5cm,right=2.5cm, bottom=2.5cm]{geometry}
\usepackage[usenames,dvipsnames]{xcolor}

\usepackage[superscript]{cite}
\usepackage[ddmmyyyy,hhmmss]{datetime}
\usepackage{bm}
\usepackage{amsmath}
\usepackage{multirow}

\begin{document}

\begin{center}
\textbf{\Large Fast and slow earthquakes emerge due to fault geometrical complexity} \\[16pt]
Pierre Romanet$^{1,2}$, Harsha S. Bhat$^{2}$, Romain Jolivet$^{2}$, Ra\'{u}l Madariaga$^{2}$\\[16pt]
\end{center}

\begin{enumerate}
\item{Institut de Physique du Globe de Paris, CNRS-UMR 7154, Sorbonne Paris Cit\'{e}, Paris 75005, France}
\item{Laboratoire de G\'{e}ologie, \'{E}cole Normale Sup\'{e}rieure, CNRS-UMR 8538, PSL Research University, Paris 75005, France.}
\end{enumerate}

\begin{center}
\textbf{Re-submitted to Nature Geoscience}\\
%Prepared on \today\ at \currenttime
\end{center}

\baselineskip18pt

\textbf{Active faults release elastic strain energy via a whole continuum of modes of slip, ranging from devastating earthquakes to Slow Slip Events and persistent creep\cite{peng2010}. Understanding the mechanisms controlling the occurrence of rapid, dynamic slip radiating seismic waves (i.e. earthquakes) or slow, silent slip (i.e. SSEs) is a fundamental point in the estimation of seismic hazard along subduction zones\cite{obara2016}. On top of showing slower rupture propagation velocity than earthquakes, SSEs exhibit different scaling relationships\cite{ide2007}, which could reflect either different physical mechanisms or an intriguing lack of observations\cite{peng2010}. Like earthquakes, SSEs are bound to occur along unstable portions of active faults, raising the question of the physical control of the mode of slip (seismic or aseismic) along these sections. Here, we use the numerical implementation of a simple rate-weakening fault model to explain the spontaneous occurrence, the characteristics and the scaling relationship of SSEs and earthquakes. We show that the simplest of fault geometrical complexities with uniform friction properties can reproduce slow and fast earthquakes without appealing to complex rheologies or mechanisms. Our model helps resolve many of the existing paradoxes between observations and physical models of earthquakes and SSEs.}   

Since their discovery in the late nineties, Slow-Slip Events (SSE) have been widely observed along various subduction zones (Central Ecuador\cite{vallee2013b}, Bungo Channel\cite{hirose1999}, Guerrero\cite{lowry2001}, Cascadia\cite{dragert2001,rogers2003}, Hikurangi\cite{douglas2005}, Northern Chile\cite{ruiz2014} and others). The discovery of SSEs mainly came from the development and the installation of networks of permanent GPS stations around subduction zones. Although GPS is still nowadays the main SSE detection tool, new observations now allow for the detection of slow-slip, like INSAR\cite{rousset2016,jolivet2013}, networks of sea-bottom pressure gauge\cite{ito2013,wallace2016} or, indirectly, via the migration of microseismicity, repeating earthquakes and tremors\cite{igarashi2003,kato2012}, thus increasing significantly the probability of their detection. 

SSEs, like earthquakes, correspond to an accelerating slip front propagating along a fault. However, unlike earthquakes, SSEs do not radiate any detectable seismic waves and are hence sometimes nicknamed ``silent events''. Until the discovery of SSEs, it was thought that only earthquakes release the accumulated strain energy along a fault. Since SSEs also contribute to this release of energy, they should play an important role in the estimation of seismic hazard along subduction zones\cite{obara2016}. In addition, SSEs exhibit very specific characteristics. Their propagation speed along the fault (about 0.5 km/h in Cascadia\cite{dragert2004} to about 1 km/day in Mexico \cite{franco2005}) contrasts with the rupture propagation speed of earthquakes (at about 3 km/s). The slip velocity of SSEs (from about 1mm/yr in the Bungo Channel, Japan to about 1 m/year in Cascadia) is around one or two orders of magnitude greater than plate convergence rates but orders of magnitude smaller than earthquakes slip rates (of the order of 1m/s)\cite{schwartz2007}. 

Although the exact influence of SSEs in the seismic cycle is not yet fully understood, they seem closely related to earthquakes. Several seismic and geodetic observations suggest that SSEs may have happened just before and in regions overlapping with earthquakes. The 2011 $\mathrm{M_{w}}$ 9.0 Tohoku-Oki event and the 2014 $\mathrm{M_{w}}$ 8.1 Iquique event are two examples in subduction zones were a SSE apparently occurred just before the earthquake, within a region overlapping with the area where seismic slip nucleated \cite{kato2012,brodsky2014,ruiz2014, mavrommatis2015}. More recently, geodetic evidence of a large SSE triggering an earthquake was pointed out in the Guerrero subduction zone\cite{radiguet2016}. There are also suggestions that SSEs may be triggered by earthquakes either by stress-waves and/or static stress transfer\cite{itaba2011,zigone2012,kato2014b,wallace2017}. On the other hand some SSEs occur without an accompanying large earthquake as in the Cascadia subduction zone, where SSEs occur periodically\cite{rogers2003}, or in the Hikurangi subduction zone\cite{wallace2016}. Yet, despite numerous observations and quantifications, the underlying physical mechanism driving SSEs remains largely unexplained. All SSEs have the same sense of slip as earthquakes, i.e. opposite to the plate convergence direction, and are accompanied by a positive stress drop which corresponds to a reduction in the accumulated strain energy. In the absence of external forcing mechanism, this necessitates SSEs to occur in a strength weakening region which is also prone to rupture as a fast dynamic event. These observations, put together, raise the first question. \textit{What physical mechanism explains slow-slip and fast, dynamic earthquakes occurring  under similar frictional boundary conditions along active faults?}
 
Furthermore, earthquakes and SSEs seem to follow different scaling laws\cite{ide2007}. The seismic moment of earthquakes scales with the cube of their duration (${M} \propto {T}^3$) whereas the corresponding moment of SSEs is proportional to their duration (${M} \propto {T}$), raising the second question. \textit{Is such different scaling a general feature of earthquakes and SSEs, highlighting different physical mechanisms? Or, is the gap in between these scaling laws simply resulting from a lack of observation\cite{ide2008,peng2010}?} We address the above questions using physics-based numerical modeling of active faults governed by rate-and-state friction\cite{dieterich1978} and develop a unified framework that addresses all the observations about earthquakes and SSEs mentioned above. 

SSEs were first discovered to emerge spontaneously from numerical models in the rate-and-state framework for the modelling of subduction zones\cite{liu2005,liu2007}. In this framework, fault areas with weakening properties will preferentially host seismic slip (i.e. earthquakes) while strengthening regions will host stable continuous creep or post-seismic slip. Numerical experiments and theoretical works have shown that the main physical control on the emergence of SSEs in models is how the characteristic length of a weakening patch compares to the specific nucleation length scale\cite{ruina1983,rice1983,dieterich1992,rubin2005}. If the length of a fault patch is large compared to the nucleation length scale, earthquakes have enough room to grow and become dynamic, so this fault patch will generate only dynamic, seismic events. If the length of the fault is small compared to this length scale, earthquakes can never grow large enough to become dynamic or no events will occur at all (i.e. permanent creep). It is therefore necessary, under this framework, to tune for the right fault length compared to the nucleation length scale to allow modelling of both slow and fast ruptures. Given the observed spatial size over which some SSEs propagate i.e. on the order of tens of kilometres, this would lead to unrealistically large nucleation sizes, preventing the occurrence of any earthquakes. A possible explanation for such large nucleation lengths could be the presence of high-pressure pore fluids released during metamorphic dehydration reactions. However it has been shown recently that regions of high fluid pressure and slow slip events do not always overlap along all the subduction zones\cite{saffer2015}. One solution to overcome this issue is to appeal to other competing frictional mechanisms like dilatant-strengthening\cite{segall1995,rubin2008,segall2010a} with or without thermal-pressurization\cite{segall2012}. Although we do not include these additional frictional mechanisms, we acknowledge that it would broaden the range over which we are able to observe slow-slip.

Our work here differs from the above line of reasoning as we do not impose any lateral variation in the rheological properties of the fault. Our aim is to introduce no a priori complexity in initial and boundary conditions and let the variety of modes of slip emerge spontaneously. As the above models suggest, a set of competing mechanisms are required for slow-slip and earthquakes to coexist. One ubiquitous feature that is often left aside for computational reasons is the geometric complexity of active faults. Indeed, faults are rarely planar over length scales of tens of kilometres and in fact, fault segmentation and geometric complexity are visible at multiple scales\cite{candela2012}. This non-planarity of faults introduces a natural stress based interaction between faults that can encourage or inhibit slow-slip. Several lines of evidence suggest that geometric complexity is a viable candidate to explain the various observed slip dynamics. Aseismic slip has been observed with earthquake swarms in the northern Appenines (Italy) along splay faults \cite{gualandi2017b}. It has been detected along the Haiyuan fault (China)\cite{jolivet2013}, the North Anatolian Fault \cite{rousset2016,bilham2016} or, in earlier publications, along the San Andreas Fault \cite{murray2005}. SSE's have been observed in the very shallow part of subduction zones, as in Hikurangi \cite{wallace2016}, Nankai \cite{araki2017} among others. A common ingredient of all these different seismo-tectonic contexts is the geometrical complexity of faults across scales.

\begin{figure}[!t]
%183mm 120mm 89mm%
\centering
\includegraphics[width=183mm]{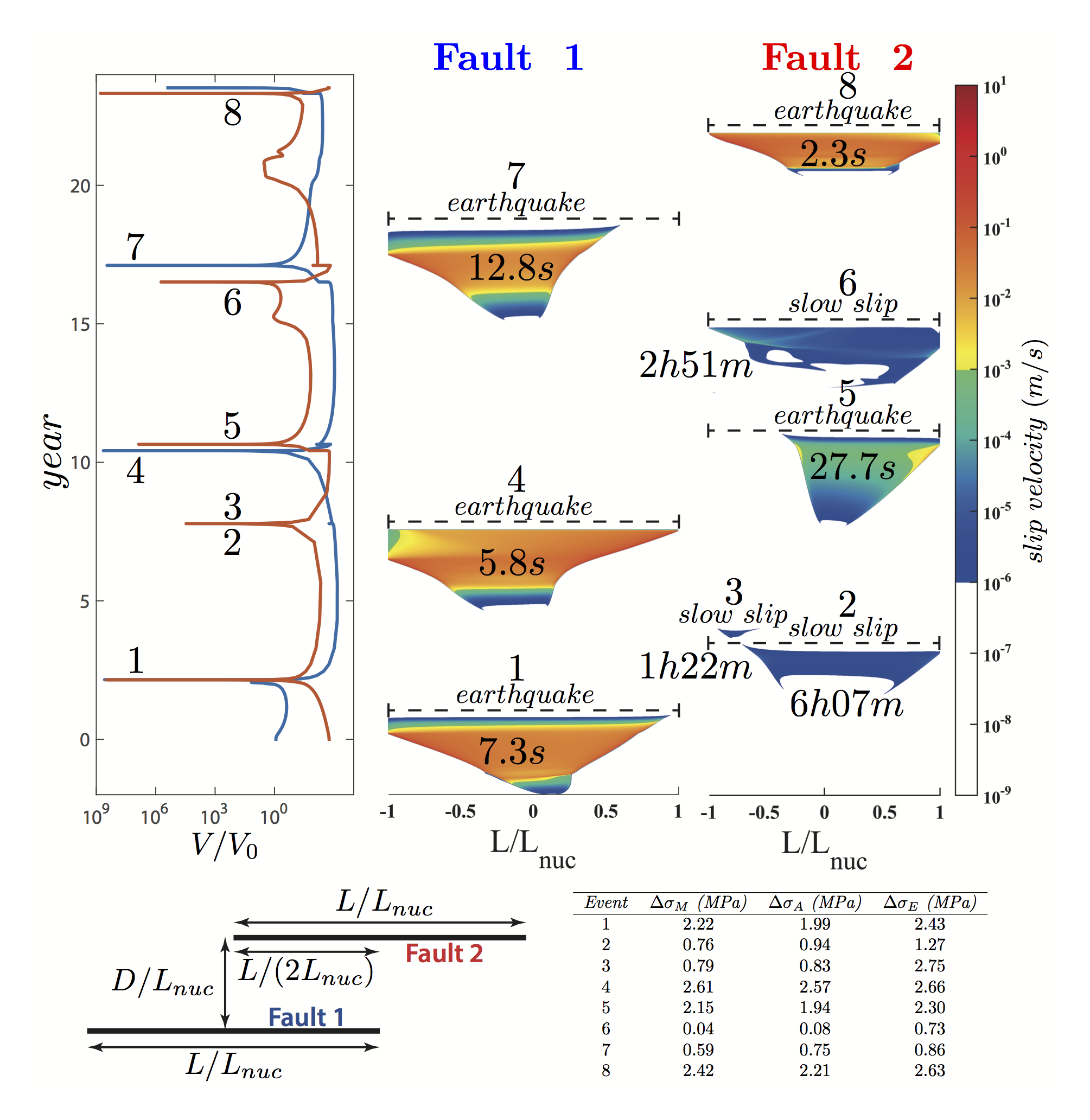}
\caption{Example of a calculation that gives rise to complex slip behaviour on faults. Here $L/L_{nuc} = 2$, $D/L_{nuc} = 0.1$ and $a/b = 0.9$. To avoid any artefact from initial conditions, the first 10 events of the simulation shown were removed.  Left panel shows the maximum slip velocity for fault 1 (blue) and fault 2 (red). Right panel represents the space-time evolution of slip velocity on the faults. The highlighted duration of events corresponds respectively for earthquakes and slow events to the time when the slip velocity exceeds 1mm/s-1$\mu$m/s for the first time to the time when it decelerates below 1mm/s-1$\mu$m/s. Bottom panel gives the geometry used for this example. Events 2,3 and 6 are slow-slip events. Events 1, 4, 5, 7 and 8 are earthquakes. Event 5 and 7 are small earthquakes that did not rupture the entire fault. Event 1 and 7 clearly show afterslip contrary to events 4 and 8. The table lists the seismological ($\Delta\sigma_{M}$), spatially averaged ($\Delta\sigma_{A}$) and slip averaged ($\Delta\sigma_{E}$) stress drops for the events.}
\label{example}
\end{figure}

In the simplest ``conceptual'' model where faults are geometrically complex, we consider two overlapping faults of the same length $L$ that interact with each other (see geometry in Fig. \ref{example}). \textit{This geometry was chosen to illustrate the effect of complex stress interaction between neighbouring faults and is in no way supposed to be interpreted as the only geometrical configuration of faults in a fault network.} Friction on both faults is controlled by rate-and-state friction with ageing state evolution. Frictional resistance decreases with increasing slip rate and is spatially uniform i.e. the fault is rate-weakening. Loading is imposed using a constant rate of shear stress increase on the fault. We model elastic interactions using out-of-plane static stress interactions with radiation damping approximation\cite{rice1993}. The computation of static stress interactions is accelerated using Fast Multipole Method, allowing us to compute all stages of the earthquake cycle in a tractable computational time\cite{greengard1987,carrier1988} (See Methods section for more details). 

The choice of such geometry brings realistic perturbations in stress along the fault leading to the emergence of a wide variety of modes of slip. Fig. \ref{example} illustrates the complexity that emerges by only appealing to stress perturbations from a neighbouring fault. We see regular earthquakes with a clear nucleation, dynamic and afterslip phases. These dynamic events happen without any evident periodicity. The novelty of this work is that slow-slip events also emerge spontaneously on the same parts of the fault (whose lengths are larger than the nucleation length) that hosted dynamic earthquakes. Without the introduction of a second fault, and its associated stress perturbations, the fault behaves like a simple spring-slider system with weakening properties, with similar earthquakes happening periodically (see Supp. Mat.).
 
\begin{figure}[!t]
\centering
\includegraphics[width=183mm]{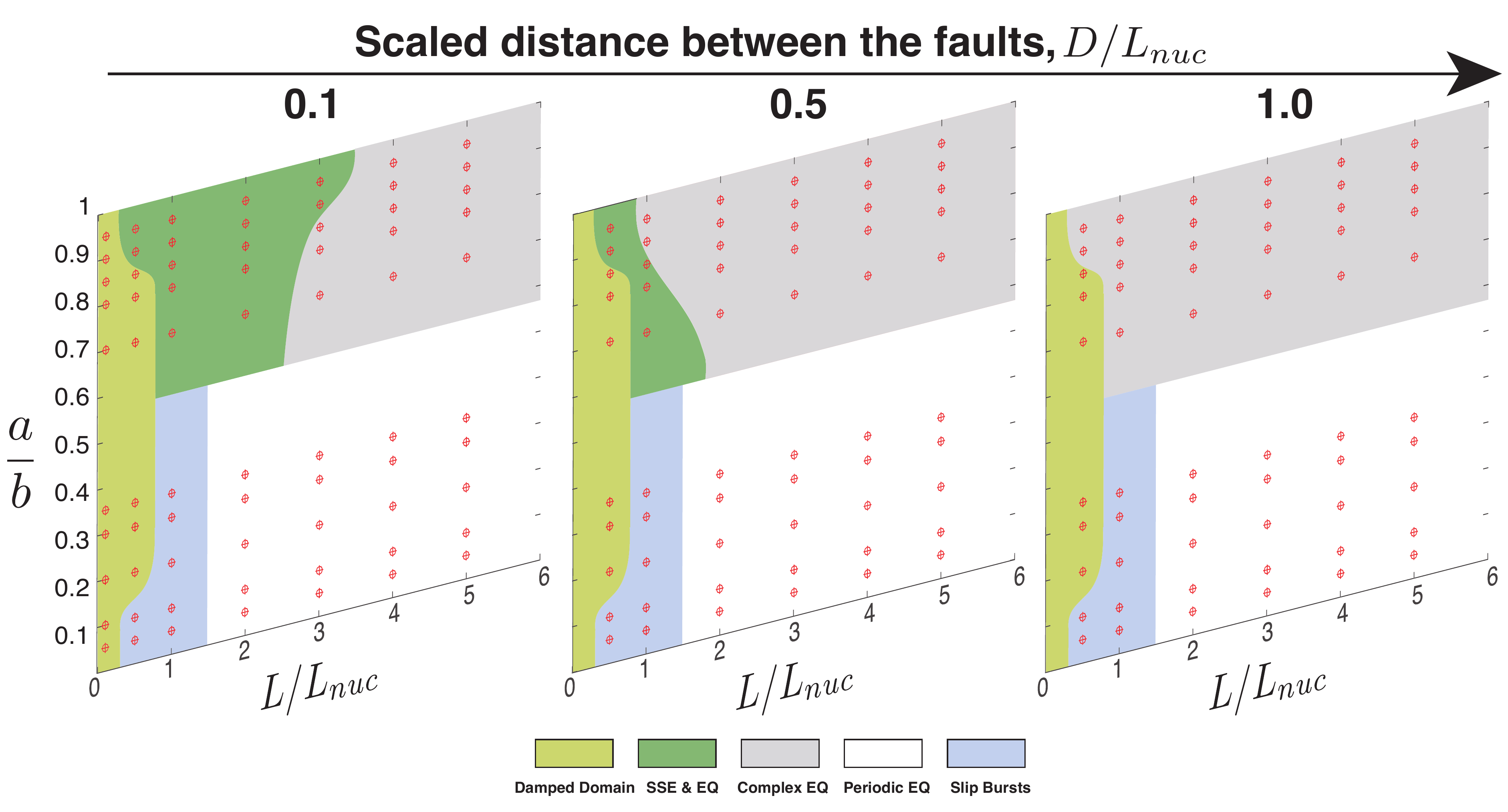}
\caption{Phase diagram showing the evolution of mode of slip along the 2 fault system given the distance between the faults. \emph{Damped domain} is a domain within witch the fault experiences no event at all. \emph{SSE \& EQ} is the domain of coexistence of both slow events and earthquakes. \emph{Complex EQ} is a domain within witch we get only earthquakes but with spatio-temporal complexities. \emph{Periodic EQ} is a domain within witch earthquakes are periodically rupturing the entire fault. And finally, \emph{Slip Bursts} is a domain within witch the entire fault is destabilized at the same time, there is no propagation of the rupture. This corresponds for small faults compared to the nucleation lenghscale and small $a/b$. This domain is called the no-healing regime\cite{rubin2005}. This figure includes a larger dataset than the one used in the study.}
\label{phaseDiagram3D}
\end{figure}

To better understand the role of multi-fault interactions on slow and fast dynamics we explored the influence of the distance between faults, $D$, the length of the faults, $L$, and the ratio of the constitutive frictional parameters, $a/b$. For rate-weakening faults, $a/b$ ranges between 0 and 1. Because of the importance of the nucleation length scale $L_{nuc}$ in this problem, all geometrical parameter are non-dimensionalized by $L_{nuc}$, 
\begin{equation}
	L_{nuc} = \frac{2}{\pi}\frac{\mu D_c}{\sigma_{n} b(1-a/b)^2} \mathrm{ ~~~  ; ~~~  } a/b \to 1
\label{nucleation}	
\end{equation}
where, $a$ and $b$ are rate-and-state constitutive friction parameters, $D_c$ is the characteristic slip distance, $\mu$ is the shear modulus of the medium and $\sigma_n$ the normal stress acting on the fault\cite{rubin2005,viesca2016a}. This formulation provides good insights on the nucleation phase of earthquakes along a fault that is mildly rate-weakening ($a/b\to 1$). For computational reasons, we restrict our experiments to fault lengths $L/L_{nuc}\in \{1,2,3,4 \}$, in order to focus on the statistics of slow and rapid slip. Our parameter space includes also distances between faults $D/L_{nuc} \in \{ 0.1,0.5, 1, 2, 3, 4\}$, and constitutive parameters $a/b \in \{0.7,0.8, 0.85, 0.90, 0.95 \}$. For illustrative purposes we provide a table of dimensional values of $L$ and $D$ in the supplementary section. The smallest faults are 200\,m long separated by distance of 21\,m. The largest faults are about 20\,km long separated by a distance of about 2\,km. In fact, it is possible to distinguish between different domains of behavior, that mainly depend on $a/b$, $L/L_{nuc}$ and the scaled distance between the faults $D/L_{nuc}$. The domain where both slow and fast earthquake coexist, shrinks when the distance between the faults is increased (Fig. \ref{phaseDiagram3D}). 

For each of the parameters identified above, we spin up the model, and allow the faults to undergo multiple earthquake cycles before measuring the slip and rupture velocity of each slow and dynamic event. We identify SSEs and earthquakes based on their slip and rupture velocity. SSEs are events with a slip velocity $V$ in the range of 1$\mathrm{\mu}$m/s to 1 mm/s and a rupture velocity $V_{rup}$ lower than $0.001c_{s}$, where $c_{s}$ is the shear wave speed. Earthquakes are events with a slip velocity greater than 1 mm/s and a rupture velocity greater than $0.001c_{s}$. We purposefully chose a relatively small threshold value for rupture velocity, because quasi-dynamic simulations lead to much slower rupture velocity than dynamic simulations\cite{thomas2009}. As our faults are one dimensional, we define the equivalent moment for a seismic or aseismic event as $M = \mu \bar{D} L_{rup} \times 1 km$, where $L_{rup}$ is the total length of the fault that slipped during an event (SSE or earthquake) and $\bar{D}$ is the slip averaged over the length $L_{rup}$. For earthquakes, we compute separately the seismic moment during the nucleation phase and the dynamic phase. For SSEs, moment accounts for the entire duration when the slip velocity exceeds $1$ $\mu$m/s. We obtained about 3000 individual earthquakes and about 500 SSEs in our calculations when the faults hosted both earthquakes and SSEs.

\begin{figure}[!t]
\centering
\includegraphics[width=100mm]{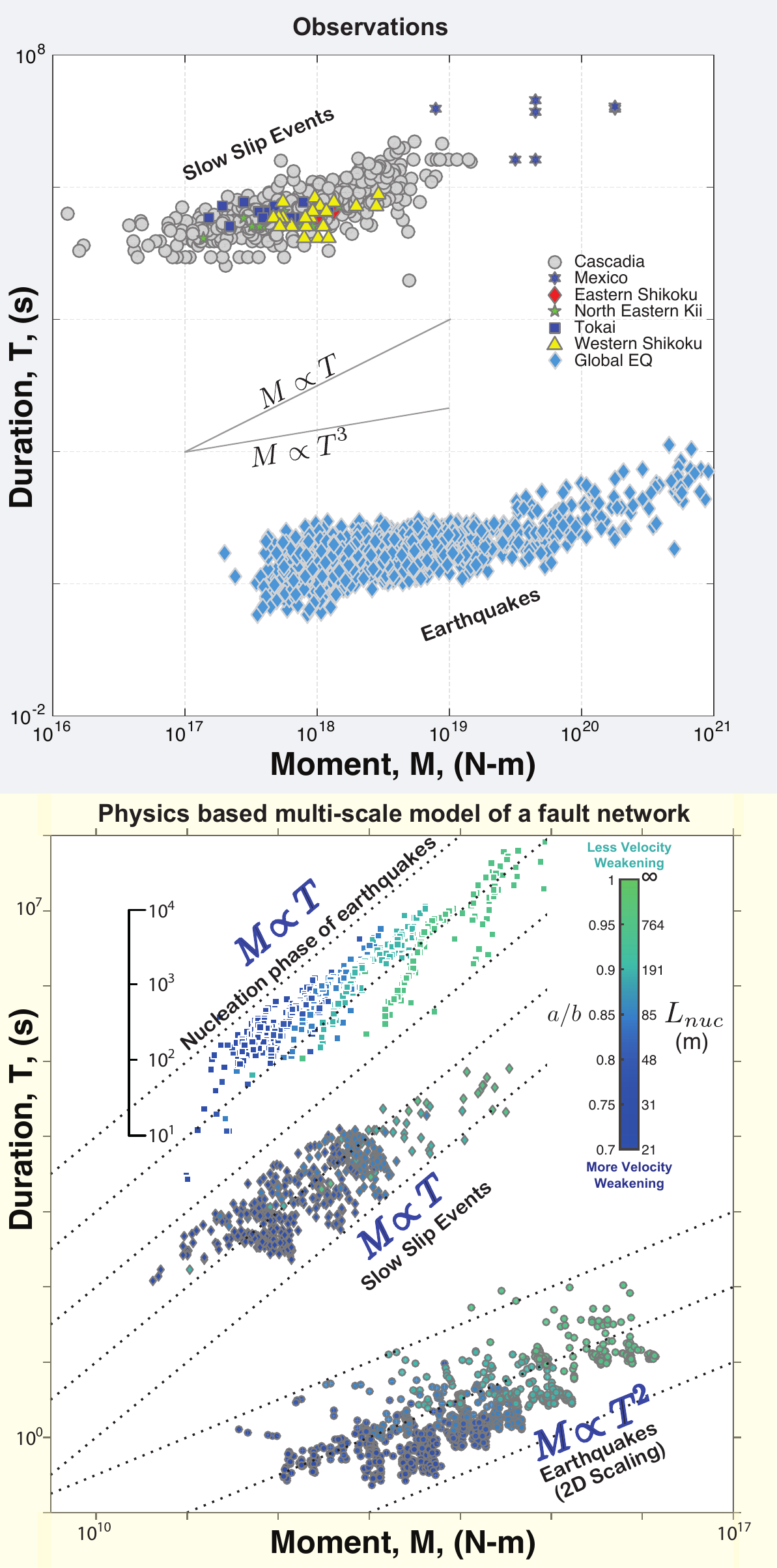}
\caption{Comparison of the scaling law for observational data\cite{sekine2010,gao2012,gomberg2016} (top panel) and from our all our calculations (bottom panel). We only used the seismic moment of the dynamic part of an earthquake.} 
\label{scaling}
\end{figure}

We find that the moment of both seismic and aseismic events modelled by rate and state friction law follows the same scaling as for events in nature\cite{ide2007,peng2010} (Fig. \ref{scaling}). Because we conducted our calculations in 2D, the moment of a dynamic event scales with its duration squared: $M  \propto T^2$. The moment of our simulated events clearly depends on the ratio of constitutive parameters $a/b$. Since the nucleation length $L_{nuc}$ increases with $a/b$ and since we compare models with non-dimensionalised fault length, the real length of the fault, $L$, also increases when $a/b {\to} 1$, leading to bigger moment release and longer duration for events. To verify the robustness of this scaling law, we changed the maximum slip velocity criteria used to distinguish SSEs and earthquakes by one order of magnitude. This does not change the observed scaling. 

\begin{figure}[!t]
\centering
\includegraphics[width=120mm]{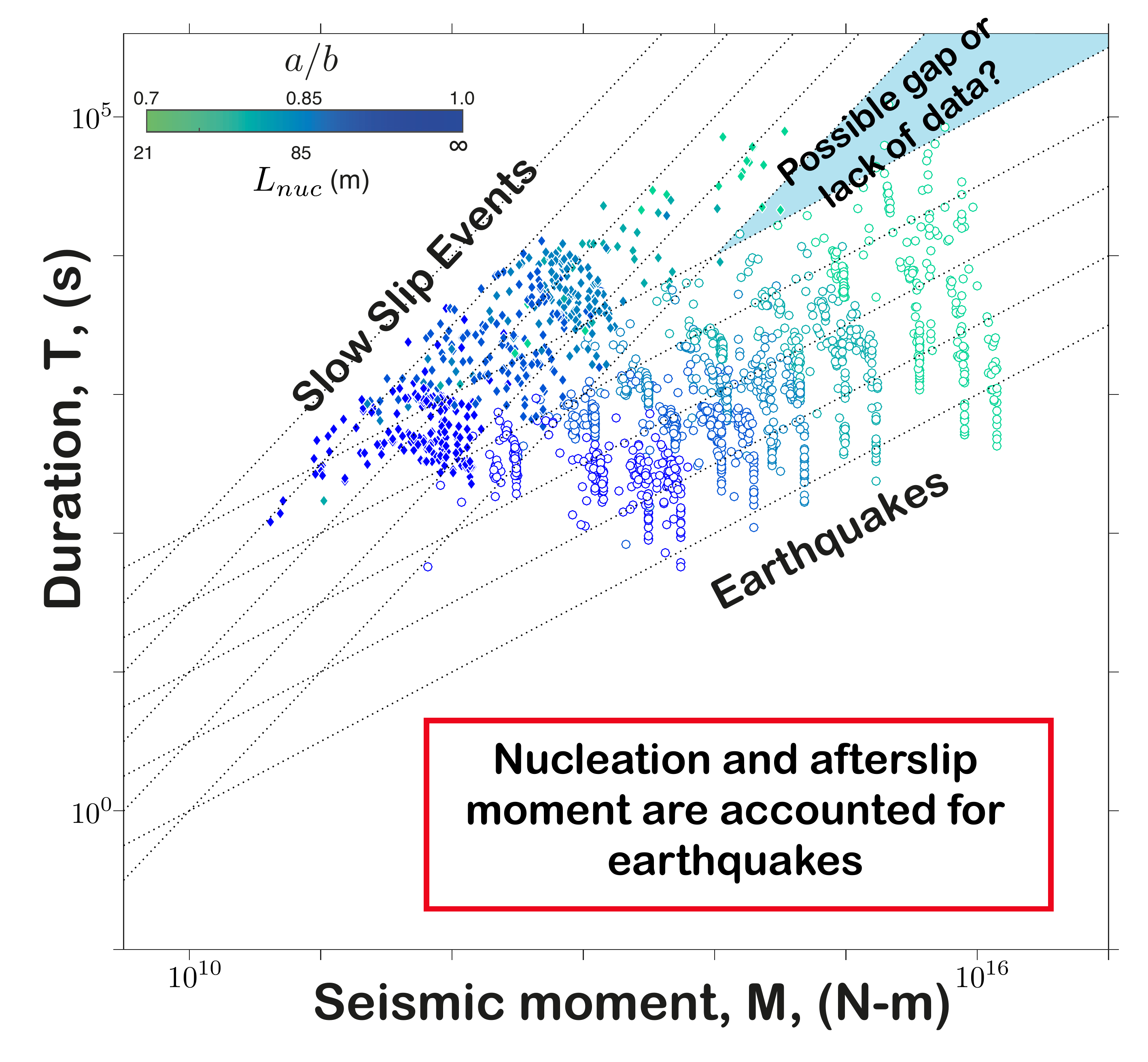}
\caption{Same as Fig. \ref{scaling} but the moment of the earthquakes includes the nucleation and afterslip phases.}
\label{explanation2}
\end{figure}

The scaling emerges naturally from our conceptual model of fault geometric complexity, without imposing any complexity in the spatial variation of frictional properties. However, we do not preclude the possibility that other models that have produced SSE's and earthquakes also reproduce such scaling laws. Another interesting feature that emerges from our calculations is that the moment of the nucleation phase of earthquakes also follows the same linear scaling with duration as slow-slip events. However, this similarity in scaling may disappear in 3D. We also notice that by adding the nucleation and after-slip moment of earthquakes, the clear scaling distinction between earthquakes and SSEs starts vanishing and a continuum between the two modes of slip may be considered (Fig. \ref{explanation2}).

\begin{figure}[!t]
\centering
\includegraphics[width=120mm]{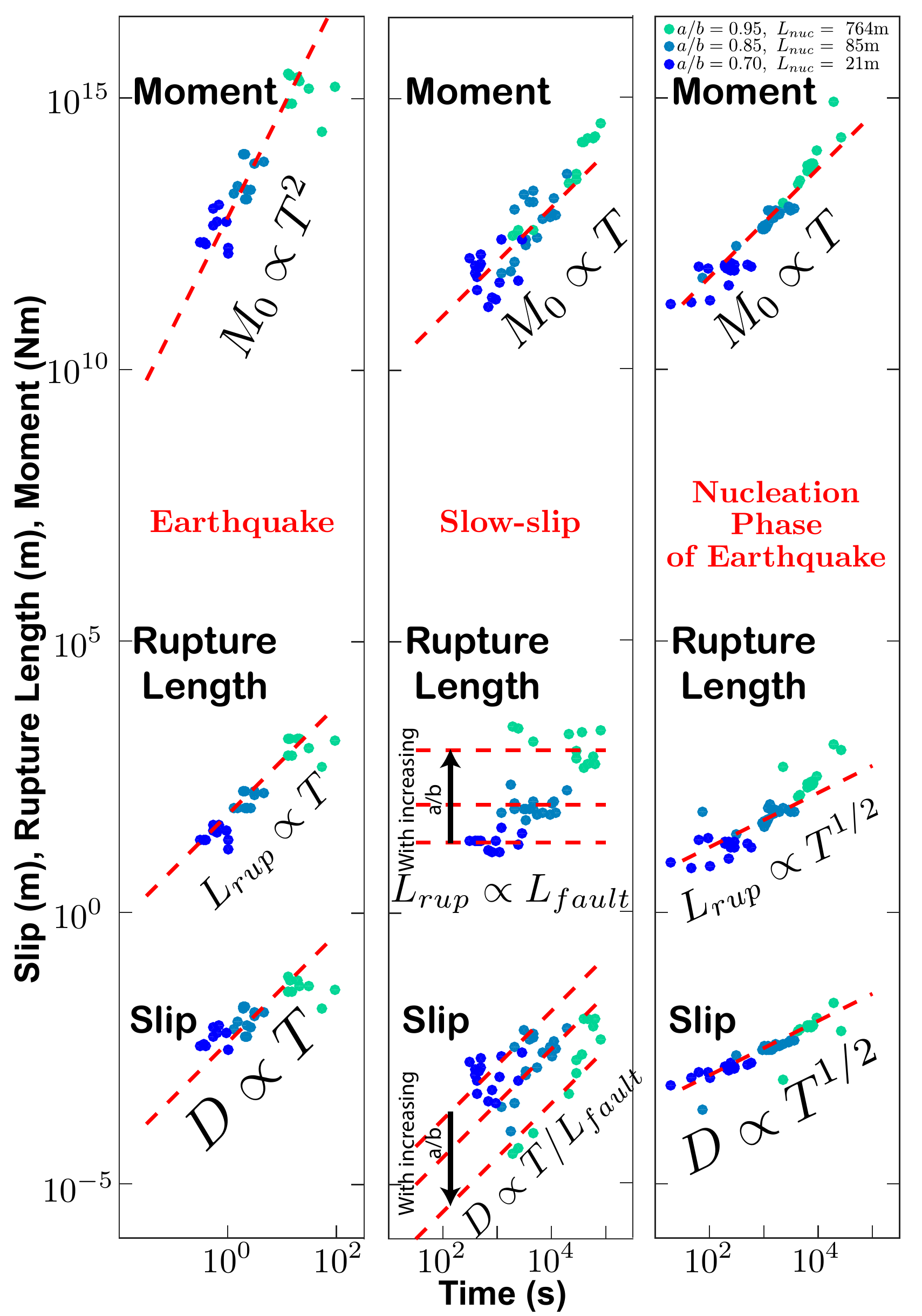}
\caption{Final moment, slip and rupture length with time for slow-slip events, earthquakes and nucleation phase of earthquakes.}
\label{explanation}
\end{figure}

The temporal evolution of rupture length and slip for each event provides hints about the relative scaling between SSEs and earthquakes (Fig. \ref{explanation}). For earthquakes, the average growth of both rupture length and slip are linear with event duration, independent of $a/b$, hence independent of the actual length of the fault as we non-dimensionalised length scales by $L_{nuc}$. As a consequence, seismic moment grows quadratically with event duration. In other words, earthquakes propagate as an expanding crack: slip and rupture length are proportional to each other. 
For SSEs, however, the temporal evolution of slip and rupture length show a clear dependence on the fault length. For a given $a/b$, final rupture length is constant i.e. it is independent of event duration. However, slip grows linearly with duration. If we now increase the fault length (i.e. increase $a/b$), the accumulated slip decreases (compared to the low $a/b$ case) while the final rupture length increases (see arrows in slow-slip panel in Fig. \ref{explanation}). These two effects exactly counterbalance each other, such that the final moment scales linearly with duration and is independent of fault length (i.e. for different $a/b$). This highlights an interesting fact that SSEs are not necessarily self-similar at least in our calculations. %For earthquakes we needed to change the fault nucleation length by nearly two orders of magnitude to span around 5 orders of magnitude of moment and two orders of magnitude of duration. For slow-slip, there is no need to change the fault length to span several orders of magnitude for moment and duration: for example the simulation for $L_{nuc}=764$ m already spans 2 orders of magnitude in duration and 3 orders of magnitude of moment.
Another interesting scaling that emerges is in the evolution of the moment of the nucleation phase with duration. It is also linear as for SSEs. The evolution of slip and rupture length for the nucleation phase is scale independent contrary to SSEs. Slip and final rupture length for nucleation phases evolve, individually, with the square root of the event duration.

Interestingly, the static stress drop of both types of slip events i.e. SSEs and earthquakes are comparable (Fig. \ref{stressdrop}). We evaluated this parameter in three different ways \cite{noda2013b}(See supplementary materials for more details). Regardless of the method, the stress drops of SSEs and earthquakes are of similar order of magnitude. Earthquake stress drops are, on an average, about twice as large as those for SSEs. Also, as expected, the stress drop scales with the moment of individual earthquakes and SSEs. Such observation emphasises the relative importance of slow events in the stress/energy budget of active faults.

Our work here suggests that slow-slip event dynamics may be controlled by fault geometrical complexities just as it has been shown to control the dynamics of ordinary earthquakes\cite{lay1981}. Unlike the current \textit{planar fault} asperity based rate-and-state models (with rate-strengthening and rate-weakening patches), the faults in our model are uniformly rate-weakening. Thus, the same segment of a fault can host both slow-slip events and earthquakes (events 5,6 and 8 in Fig. \ref{example}). This is not possible in the asperity based models since a large rate-weakening asperity ($L > L_{nuc}$) will always rupture seismically and a small rate-weakening asperity ($L\sim L_{nuc}$) will sometimes lead to aborted nucleation of dynamic events\cite{veedu2016}. However, as we have shown in Fig. \ref{explanation}, the rupture length and slip during the nucleation phase follow different scaling behaviour as opposed to slow-slip events. 

Numerous natural observations like occurrences of unexpected spontaneous slow-slip events\cite{rousset2016,wallace2016}, and more generally all SSEs, cannot be explained by the current asperity-based rate-and-state models without appealing to competing mechanisms. Within our framework, there is no need to exclusively invoke more complex weakening processes in rate-strengthening zones, like thermal pressurization, to explain shallow slip or to ad-hoc tuning of parameters. The greatest strength of the asperity based model is to explain the occurrence of afterslip by the relaxation of a large stress perturbation in a rate-strengthening region of a fault\cite{perfettini2010}. Our model shows that a fault segment next to a rupture zone can undergo aseismic slip as it would do in the case of afterslip (Fig. \ref{example}). Now that we have shown that complex stress perturbations, like those induced by complex fault geometry, lead to the emergence of a whole complexity of modes of slip, it would not be unsafe to imagine active faults with only weakening properties, either spatially homogeneous or heterogeneous. We do not exclude the possibility that natural faults do obey an asperity based model. However, a unified model that explains all the observations has to invariably account for geometric segmentation and/or the non-planar nature of active faults. Geometry is a first-order and documented feature that results in a spatio-temporally inhomogenous stress accumulation rate\cite{mitsui2006,matsuzawa2013,li2016} and it affects strongly the modes of slip throughout the cycle.

%For the range of parameters we used, $L_{nuc}$ varies from few tens of meters to a about a kilometer because we assume a high normal stress of 100 MPa. Since anti-plane elasticity results in no change in normal stress due to slip on a fault, we can scale the nucleation length and the size of the ruptures for realistic values of normal stress on subduction zone interfaces. Normal stress is presumed to range from about 10 MPa at seismogenic depths to about 1 MPa down to depths of around 50 km\cite{segall2012}. %
%As the nucleation length is inversely proportional to the normal stress the corresponding length scales over which we observe slow-slip and earthquakes scales up by a factor of ten (hundreds of meters to about 10 km) at seismogenic depths and by a factor of hundred deeper below (several kilometres to about tens of kilometres). 

%Our model also allows to explain the puzzling behaviour of afterslip following the Pedernales, $\mathrm{M_{w}}$7.8, 2016 earthquake \question{(J-M. Nocquet pers. comm. or paper)}. There, some afterslip occurs on a fault patch that is strongly coupled during the interseismic period and some afterslip occurs on a lowly coupled region of the megathrust that hosted a slow-slip event. Such behaviour contradicts clearly with a pre-imposed spatially variable fault rheology that would control the behaviour of the fault. Our model, on the other hand, would even allow to explain delayed triggering via a much simpler mechanism than the complex non-linear interaction of seismic waves within fault gouges\cite{johnson2008}. 

In this paper, we showed that a simple, conceptual, physics based mechanical model (two interacting faults with an overlap) can produce slow-slip events and earthquakes on the same rate-weakening segment of a fault whose length is much larger than the nucleation length. We also reproduce the observed scaling law of moment with the duration of an event. This is, to our knowledge, the first time that the scaling law for slow and fast events is reproduced in the rate and state framework with uniform frictional properties. The only key ingredient needed in our model is continuous, aperiodic, stress perturbations from nearby faults. This is quite easily testable, as a `single fault' inferred from seismology or geodesy is in fact a network of faults at various length scales\cite{candela2012}. 

%%%%%%%%%%%%%%%%%%%%%%%%%%%%%%%%%%%%%%%%%%%%%%%%%%%%%%%%%%%%%%%%%%%%%%%%%%%%%%%%%%%%%%%%%%%%%%%%%%%%%%%%%%%%%%%%%%%

%%%%%%%%%%%%%%%%%%%%%%%%%%%%%%%%%%%%%%%%%%%%%%%%%%%%%%%%%%%%%%%%%%%%%%%%%%%%%%%%%%%%%%%%%%%%%%%%%%%%%%%%%%%%%%%%%%%%
\newpage
%\noindent \textbf{Supplementary Information} is available in the online version of the paper.\\[12pt]
%%%%%%%%%%%%%%%%%%%%%%%%%%%%%%%%%%%%%%%%%%%%%%%%%%%%%%%%%%%%%%%%%%%%%%%%%%%%%%%%%%%%%%%%%%%%%%%%%%%%%%%%%%%%%%%%%%%%
\noindent \textbf{Acknowledgements} Numerical computations were performed on the S-CAPAD platform, IPGP, France. P. R. and H. S. B. are grateful to Leslie Greengard and Zydrunas Gimbutas for the FMMLIB2D library. This article benefited from discussions with Robert Viesca and Pierre Dublanchet. P. R. acknowledges the GPX program, funded by the French National Research Agency (ANR), CGG, TOTAL and Schlumberger, for his PhD fellowship. \\[12pt]
%%%%%%%%%%%%%%%%%%%%%%%%%%%%%%%%%%%%%%%%%%%%%%%%%%%%%%%%%%%%%%%%%%%%%%%%%%%%%%%%%%%%%%%%%%%%%%%%%%%%%%%%%%%%%%%%%%%%
\noindent \textbf{Author Contributions} All authors contributed to problem design, analysis, interpretation and manuscript preparation.  \\[12pt]
%%%%%%%%%%%%%%%%%%%%%%%%%%%%%%%%%%%%%%%%%%%%%%%%%%%%%%%%%%%%%%%%%%%%%%%%%%%%%%%%%%%%%%%%%%%%%%%%%%%%%%%%%%%%%%%%%%%%
\noindent  \textbf{Author Information} %Reprints and permissions information is available at www.nature.com/reprints. 
The authors declare no competing financial interests. 
%Readers are welcome to comment on the online version of the paper. 
Correspondence and requests for materials should be addressed to P. R. (romanet@geologie.ens.fr).\\[12pt]
%%%%%%%%%%%%%%%%%%%%%%%%%%%%%%%%%%%%%%%%%%%%%%%%%%%%%%%%%%%%%%%%%%%%%%%%%%%%%%%%%%%%%%%%%%%%%%%%%%%%%%%%%%%%%%%%%%%%
%\noindent \textbf{Reviewer Information} Nature thanks XXX and the other anonymous reviewer(s) for their contribution to the peer review of this work.\\[16pt]
%%%%%%%%%%%%%%%%%%%%%%%%%%%%%%%%%%%%%%%%%%%%%%%%%%%%%%%%%%%%%%%%%%%%%%%%%%%%%%%%%%%%%%%%%%%%%%%%%%%%%%%%%%%%%%%%%%%%
\clearpage
%%%%%%%%%%%%%%%%%%%%%%%%%%%%%%%%%%%%%%%%%%%%%%%%%%%%%%%%%%%%%%%%%%%%%%%%%%%%%%%%%%%%%%%%%%%%%%%%%%%%%%%%%%%%%%%%%%%%
%%%%%%%%%%%%%%%%%%%%%%%%%%%%%%%%%%%%%%%%%%%%%%%%%%%%%%%%%%%%%%%%%%%%%%%%%%%%%%%%%%%%%%%%%%%%%%%%%%%%%%%%%%%%%%%%%%%%

%%%%%%%%%%%%%%%%%%%%%%%%%%%%%%%%%%%%%%%%%%%%%%%%%%%%%%%%%%%%%%%%%%%%%%%%%%%%%%%%%%%%%%%%%%%%%%%%%%%%%%%%%%%%%%%%%%%%
%%%%%%%%%%%%%%%%%%%%%%%%%%%%%%%%%%%%%%%%%%%%%%%%%%%%%%%%%%%%%%%%%%%%%%%%%%%%%%%%%%%%%%%%%%%%%%%%%%%%%%%%%%%%%%%%%%%%
%%%%%%%%%%%%%%%%%%%%%%%%%%%%%%%%%%%%%%%%%%%%%%%%%%%%%%%%%%%%%%%%%%%%%%%%%%%%%%%%%%%%%%%%%%%%%%%%%%%%%%%%%%%%%%%%%%%%
%%%%%%%%%%%%%%%%%%%%%%%%%%%%%%%%%%%%%%%%%%%%%%%%%%%%%%%%%%%%%%%%%%%%%%%%%%%%%%%%%%%%%%%%%%%%%%%%%%%%%%%%%%%%%%%%%%%%
\begin{figure}[!t]
\centering
\includegraphics[width=183mm]{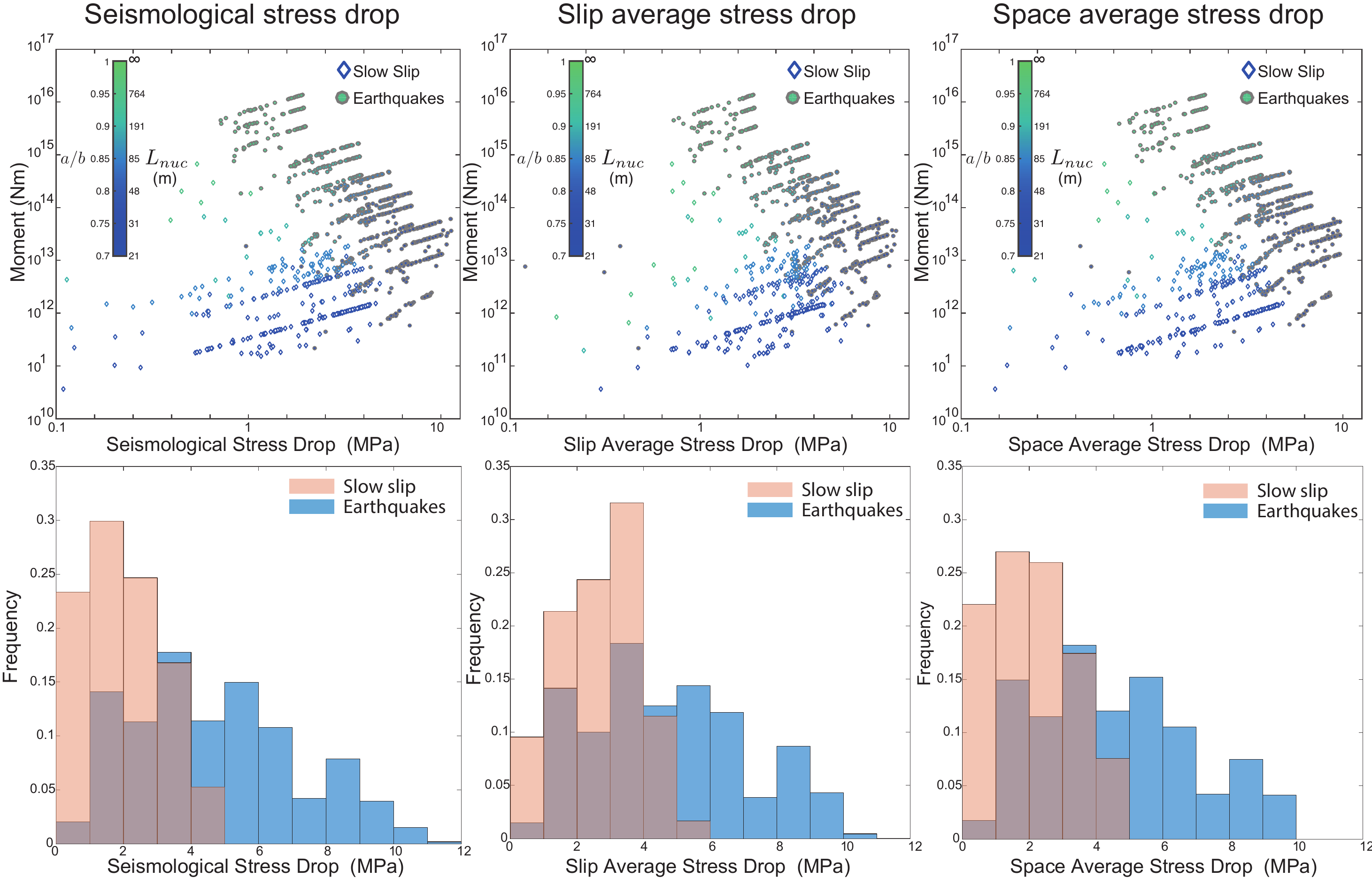}
\caption{Comparison of seismological, slip average and spatially averaged stress drops and their distributions for slow-slip events and earthquakes.}
\label{stressdrop}
\end{figure}
%%%%%%%%%%%%%%%%%%%%%%%%%%%%%%%%%%%%%%%%%%%%%%%%%%%%%%%%%%%%%%%%%%%%%%%%%%%%%%%%%%%%%%%%%%%%%%%%%%%%%%%%%%%%%%%%%%%%

%%%%%%%%%%%%%%%%%%%%%%%%%%%%%%%%%%%%%%%%%%%%%%%%%%%%%%%%%%%%%%%%%%%%%%%%%%%%%%%%%%%%%%%%%%%%%%%%%%%%%%%%%%%%%%%%%%%%
\clearpage
\noindent \textbf{\large Methods}\\
%%%%%%%%%%%%%%%%%%%%%%%%%%%%%%%%%%%%%%%%%%%%%%%%%%%%%%%%%%%%%%%%%%%%%%%%%%%%%%%%%%%%%%%%%%%%%%%%%%%%%%%%%%%%%%%%%%%%
\textbf{Modelling earthquake cycles in fault networks using the Fast Multipole Method}

Consider a network of faults in a 2D medium governed by anti-plane isotropic linear elasticity.	Let the faults be described by a curvilinear co-ordinate $s$ such that the points on the fault in the global $x_{1}-x_{2}$ co-ordinate system are given by $\bm{f}(s)=\{f_{1}(s),f_{2}(s)\}$. Let $\Delta u_{3}(s) \equiv D(s)$ be the slip distribution on these faults. The quasi-static shear traction, $T_{el}$, at a given point on the fault due to the slip distribution is given by
%%%%%%%%%%%%%%%%%%%%%%%%%%%%%%%%%%%%%%%%%%%%%%%%%%%%%%%%%%%%%%%%%%%%%%%%%%%%%%%%%%%%%%%%%%%%%%%%%%%%%%%%%%%%%%%%%%%
\begin{eqnarray}
\label{elasticstress}
T_{el}(s,t) &=& \frac{\mu}{2\pi}\int_{\Gamma} K(s,\xi) \frac{\partial D(\xi,t)}{\partial \xi} \mathrm{d}\xi
\end{eqnarray}
%%%%%%%%%%%%%%%%%%%%%%%%%%%%%%%%%%%%%%%%%%%%%%%%%%%%%%%%%%%%%%%%%%%%%%%%%%%%%%%%%%%%%%%%%%%%%%%%%%%%%%%%%%%%%%%%%%%
\begin{eqnarray}
\textrm{where~~~~~} K(s,\xi) = n_1(s)\frac{f_{2}(s)-f_2(\xi)}{r}-n_2(s)\frac{f_{1}(s)-f_1(\xi)}{r}\nonumber 
\textrm{~~~~and~~~~~} r = \sqrt{[f_{1}(s)-f_1(\xi)]^{2}+[f_{2}(s)-f_2(\xi)^{2}]}\nonumber 
\end{eqnarray}
%%%%%%%%%%%%%%%%%%%%%%%%%%%%%%%%%%%%%%%%%%%%%%%%%%%%%%%%%%%%%%%%%%%%%%%%%%%%%%%%%%%%%%%%%%%%%%%%%%%%%%%%%%%%%%%%%%%
Here $\mu$ is the shear modulus (30 GPa), $c_s$ the shear wave speed (3 km/s). $\bm{n}(s)$ is the unit normal vector to the fault(s) at point $s$ and $\Gamma$ is a contour that traces the fault(s) geometry. 
%%%%%%%%%%%%%%%%%%%%%%%%%%%%%%%%%%%%%%%%%%%%%%%%%%%%%%%%%%%%%%%%%%%%%%%%%%%%%%%%%%%%%%%%%%%%%%%%%%%%%%%%%%%%%%%%%%%
This shear traction is balanced by the strength of the fault, modelled by rate-and-state friction law with ageing state evolution:
%%%%%%%%%%%%%%%%%%%%%%%%%%%%%%%%%%%%%%%%%%%%%%%%%%%%%%%%%%%%%%%%%%%%%%%%%%%%%%%%%%%%%%%%%%%%%%%%%%%%%%%%%%%%%%%%%%%
\begin{equation}
T_f(s,t) = \left[f_0+a \log \left\{\frac{V(s,t)}{V_0} \right\}+ b \log \left\{\frac{\theta(s,t) V_0}{D_c} \right\} \right]\sigma_n 
\label{rs1}
\end{equation}
%%%%%%%%%%%%%%%%%%%%%%%%%%%%%%%%%%%%%%%%%%%%%%%%%%%%%%%%%%%%%%%%%%%%%%%%%%%%%%%%%%%%%%%%%%%%%%%%%%%%%%%%%%%%%%%%%%%
\begin{equation}
\frac{d\theta(s,t)}{dt}=1-\frac{V(s,t) \theta(s,t)}{D_c}
\label{rs2}
\end{equation}
%%%%%%%%%%%%%%%%%%%%%%%%%%%%%%%%%%%%%%%%%%%%%%%%%%%%%%%%%%%%%%%%%%%%%%%%%%%%%%%%%%%%%%%%%%%%%%%%%%%%%%%%%%%%%%%%%%%
For all our calculations we fix the weakening distance, $D_{c}$ to be 100 $\mu$m, $V_{0}= \mathrm{10^{-9}}$ m/s, $b=0.01$ and the normal stress $\sigma_{n}$ = 100 MPa.
%%%%%%%%%%%%%%%%%%%%%%%%%%%%%%%%%%%%%%%%%%%%%%%%%%%%%%%%%%%%%%%%%%%%%%%%%%%%%%%%%%%%%%%%%%%%%%%%%%%%%%%%%%%%%%%%%%%
The far-field tectonic loading rate, $\dot{T}_{\infty}$, is assumed to be constant at 0.01 Pa/s such that
%%%%%%%%%%%%%%%%%%%%%%%%%%%%%%%%%%%%%%%%%%%%%%%%%%%%%%%%%%%%%%%%%%%%%%%%%%%%%%%%%%%%%%%%%%%%%%%%%%%%%%%%%%%%%%%%%%%
\begin{equation}
T_{load}(s,t) = \dot{T}_{\infty}t
\end{equation}
%%%%%%%%%%%%%%%%%%%%%%%%%%%%%%%%%%%%%%%%%%%%%%%%%%%%%%%%%%%%%%%%%%%%%%%%%%%%%%%%%%%%%%%%%%%%%%%%%%%%%%%%%%%%%%%%%%%
By requiring the forces to balance on the fault we obtain,
%%%%%%%%%%%%%%%%%%%%%%%%%%%%%%%%%%%%%%%%%%%%%%%%%%%%%%%%%%%%%%%%%%%%%%%%%%%%%%%%%%%%%%%%%%%%%%%%%%%%%%%%%%%%%%%%%%%
\begin{equation}
T_{el} +T_{load} - \frac{\mu V}{2c_{s}}= T_f
\end{equation}
%%%%%%%%%%%%%%%%%%%%%%%%%%%%%%%%%%%%%%%%%%%%%%%%%%%%%%%%%%%%%%%%%%%%%%%%%%%%%%%%%%%%%%%%%%%%%%%%%%%%%%%%%%%%%%%%%%%
where ${\mu V}/{2c_{s}}$ is the radiation damping approximation\cite{rice1993} that accounts for instantaneous change in the shear traction due to a dynamic change in the slip (via the slip rate, $V = \partial D/\partial t$) and ignores the effect of radiated stress waves. This is the quasi-dynamic approximation to model earthquake cycles that are numerically tractable. Differentiating the above expression with time, and rearranging, leads to a set of coupled ordinary differential equations (ODE) that is then solved numerically using the Bulirsch-St\"oer\cite{bulirsch1966} adaptive time stepping method.
%%%%%%%%%%%%%%%%%%%%%%%%%%%%%%%%%%%%%%%%%%%%%%%%%%%%%%%%%%%%%%%%%%%%%%%%%%%%%%%%%%%%%%%%%%%%%%%%%%%%%%%%%%%%%%%%%%%
\begin{eqnarray}
\frac{dV(s,t)}{dt}              & = & \left[{\dot{T}_{\infty}-\left\{\frac{\mu}{2\pi}\int_{\Gamma} K(s,\xi) \frac{\partial V(\xi,t)}{\partial \xi} \mathrm{d}\xi\right\}-\frac{b\sigma_{n}}{\theta(s,t)}\left(1 - \frac{V(s,t)\theta(s,t)}{D_c}\right)}\right] \left[{\frac{\mu}{2c_s}+\frac{\sigma_{n}a}{V(s,t)}}\right]^{-1}  \\ 
\frac{d\theta(s,t)}{dt}         & = &1 - \frac{V(s,t)\theta(s,t)}{D_c}   
\end{eqnarray}
%%%%%%%%%%%%%%%%%%%%%%%%%%%%%%%%%%%%%%%%%%%%%%%%%%%%%%%%%%%%%%%%%%%%%%%%%%%%%%%%%%%%%%%%%%%%%%%%%%%%%%%%%%%%%%%%%%%

The most time consuming part is to compute the elastic stress interaction, because of the convolution term (in curly braces) in right hand part of equation (\ref{elasticstress}). This is why we appeal to the Fast Multipole Method\cite{greengard1987,carrier1988}, FMM, to accelerate the calculation of this term. Fast multipole method was initially developed in the context of N-body problems. It consists of developing the far-field multipole expansion of a group of points and then summing all those far-field contributions by the means of a Taylor series (local expansion). See the references cited for more details.

Starting from the expression for elastic stress change in equation (\ref{elasticstress}), we can obtain
%%%%%%%%%%%%%%%%%%%%%%%%%%%%%%%%%%%%%%%%%%%%%%%%%%%%%%%%%%%%%%%%%%%%%%%%%%%%%%%%%%%%%%%%%%%%%%%%%%%%%%%%%%%%%%%%%%%
\begin{eqnarray}
T_{el}(s,t) &=& -\frac{\mu}{2\pi}\bm{t}(s)\cdot \bm{{\bm{\nabla}}} \left[\int_{\Gamma} \! \frac{\partial D(\xi)}{\partial \xi} \log{|| \bm{f}(s)-\bm{f}(\xi) ||} \mathrm{d}\xi \right]\\
|| \bm{f}(s)-\bm{f}(\xi) || &= & \sqrt{[f_{1}(s)-f_1(\xi)]^{2}+[f_{2}(s)-f_2(\xi)^{2}]}
\end{eqnarray}
%%%%%%%%%%%%%%%%%%%%%%%%%%%%%%%%%%%%%%%%%%%%%%%%%%%%%%%%%%%%%%%%%%%%%%%%%%%%%%%%%%%%%%%%%%%%%%%%%%%%%%%%%%%%%%%%%%%
where $\bm{t}(s)$ is the unit tangent vector at $s$.

To evaluate this integral numerically, we assumed piece-wise constant slip on an element of length $\Delta s$. The centre of each element being given by its curvilinear coordinate $s_i$. Thus, 
%%%%%%%%%%%%%%%%%%%%%%%%%%%%%%%%%%%%%%%%%%%%%%%%%%%%%%%%%%%%%%%%%%%%%%%%%%%%%%%%%%%%%%%%%%%%%%%%%%%%%%%%%%%%%%%%%%%
\begin{equation} \label{eq:dicrete_slip}
D(s) = \sum_{elements} \left[\mathcal{H}\left(s-s_i+\frac{\Delta s}{2}\right)-\mathcal{H}\left(s-s_i-\frac{\Delta s}{2}\right) \right] D_i
\end{equation}
%%%%%%%%%%%%%%%%%%%%%%%%%%%%%%%%%%%%%%%%%%%%%%%%%%%%%%%%%%%%%%%%%%%%%%%%%%%%%%%%%%%%%%%%%%%%%%%%%%%%%%%%%%%%%%%%%%%
where $\mathcal{H}(x)$ is the Heaviside function.
%%%%%%%%%%%%%%%%%%%%%%%%%%%%%%%%%%%%%%%%%%%%%%%%%%%%%%%%%%%%%%%%%%%%%%%%%%%%%%%%%%%%%%%%%%%%%%%%%%%%%%%%%%%%%%%%%%%
\begin{figure}[!t]
\centering
\includegraphics[width=120mm]{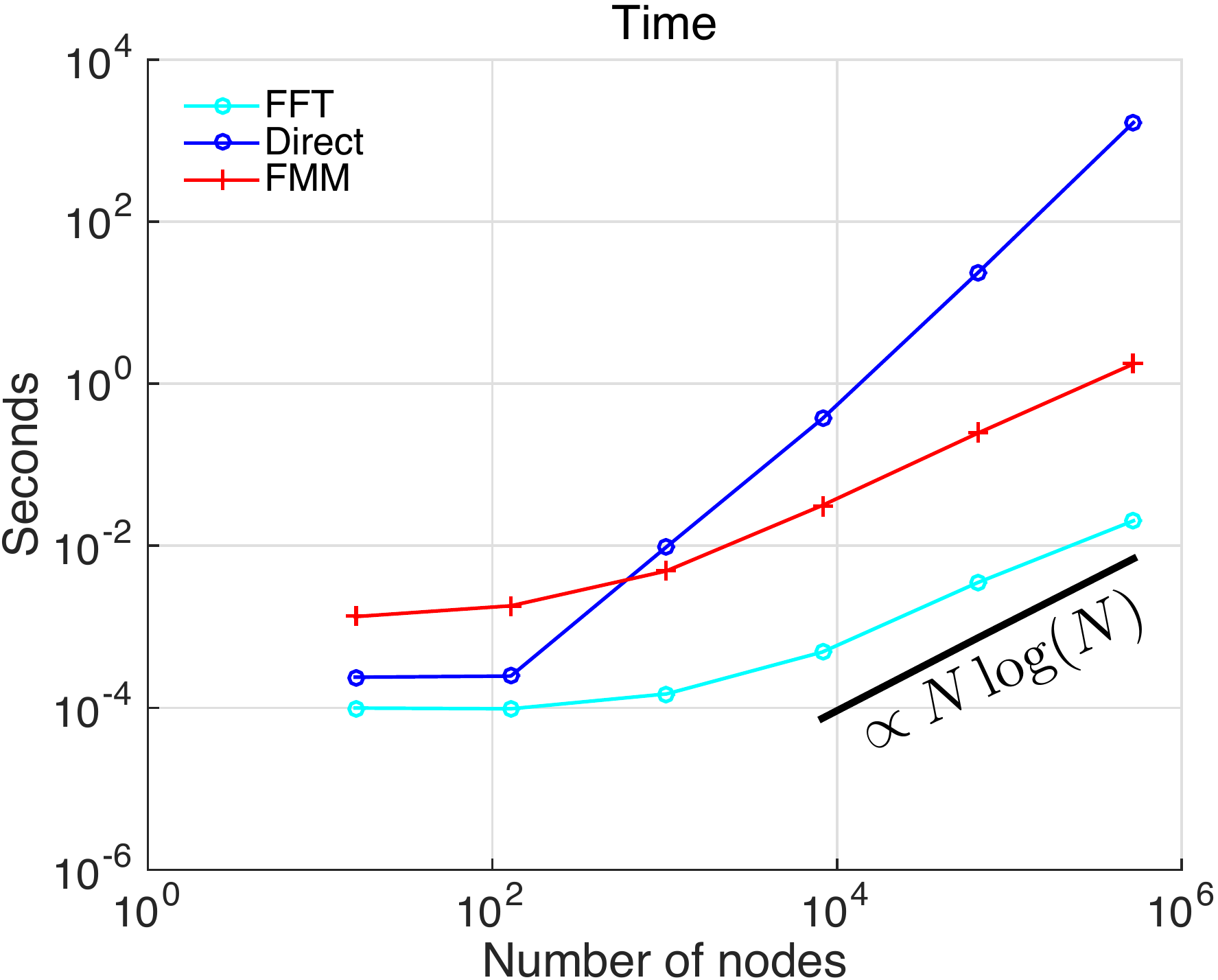}
\caption{Computational complexity of FFT, FMM and Direct methods for evaluating stress change on a fault due to a given slip distribution.}
\label{timeFMM}
\end{figure}
%%%%%%%%%%%%%%%%%%%%%%%%%%%%%%%%%%%%%%%%%%%%%%%%%%%%%%%%%%%%%%%%%%%%%%%%%%%%%%%%%%%%%%%%%%%%%%%%%%%%%%%%%%%%%%%%%%%
Thus static elastic stress can be evaluated as a sum of logarithmic functions:
%%%%%%%%%%%%%%%%%%%%%%%%%%%%%%%%%%%%%%%%%%%%%%%%%%%%%%%%%%%%%%%%%%%%%%%%%%%%%%%%%%%%%%%%%%%%%%%%%%%%%%%%%%%%%%%%%%%
\begin{eqnarray}\label{eq:discrete_traction} 
T_{el}(s,t) =\frac{\mu}{2\pi}\bm{t}(s)\cdot \bm{{\nabla}}
          \left[\sum_{elements} D_i 
          \left\{\log{|| \bm{f}(s)-\bm{f}(s_i-\Delta s/2) || -  \log || \bm{f}(s)-\bm{f}(s_i+\Delta s/2) ||} \right\}\right]
\end{eqnarray}
%%%%%%%%%%%%%%%%%%%%%%%%%%%%%%%%%%%%%%%%%%%%%%%%%%%%%%%%%%%%%%%%%%%%%%%%%%%%%%%%%%%%%%%%%%%%%%%%%%%%%%%%%%%%%%%%%%%
This formulation is now suitable to be accelerated by fast multipole method. In figure \ref{timeFMM}, we compared the acceleration provided by FMM with the direct evaluation of this summation, and the calculation through Fast Fourier Transform. For this comparison, we calculate the stress on a planar fault at each discretised point due to an elliptical distribution of slip. Although Fast Fourier transform is faster by around two orders of magnitude, it can not deal with multiple faults or complex geometries. FMM removes this restriction and is faster than the direct evaluation by 1 to 3 orders of magnitude.
%%%%%%%%%%%%%%%%%%%%%%%%%%%%%%%%%%%%%%%%%%%%%%%%%%%%%%%%%%%%%%%%%%%%%%%%%%%%%%%%%%%%%%%%%%%%%%%%%%%%%%%%%%%%%%%%%%%

\vspace{1cm}

\begin{figure}[!t]
%183mm 120mm 89mm%
\centering
\includegraphics[width=183mm]{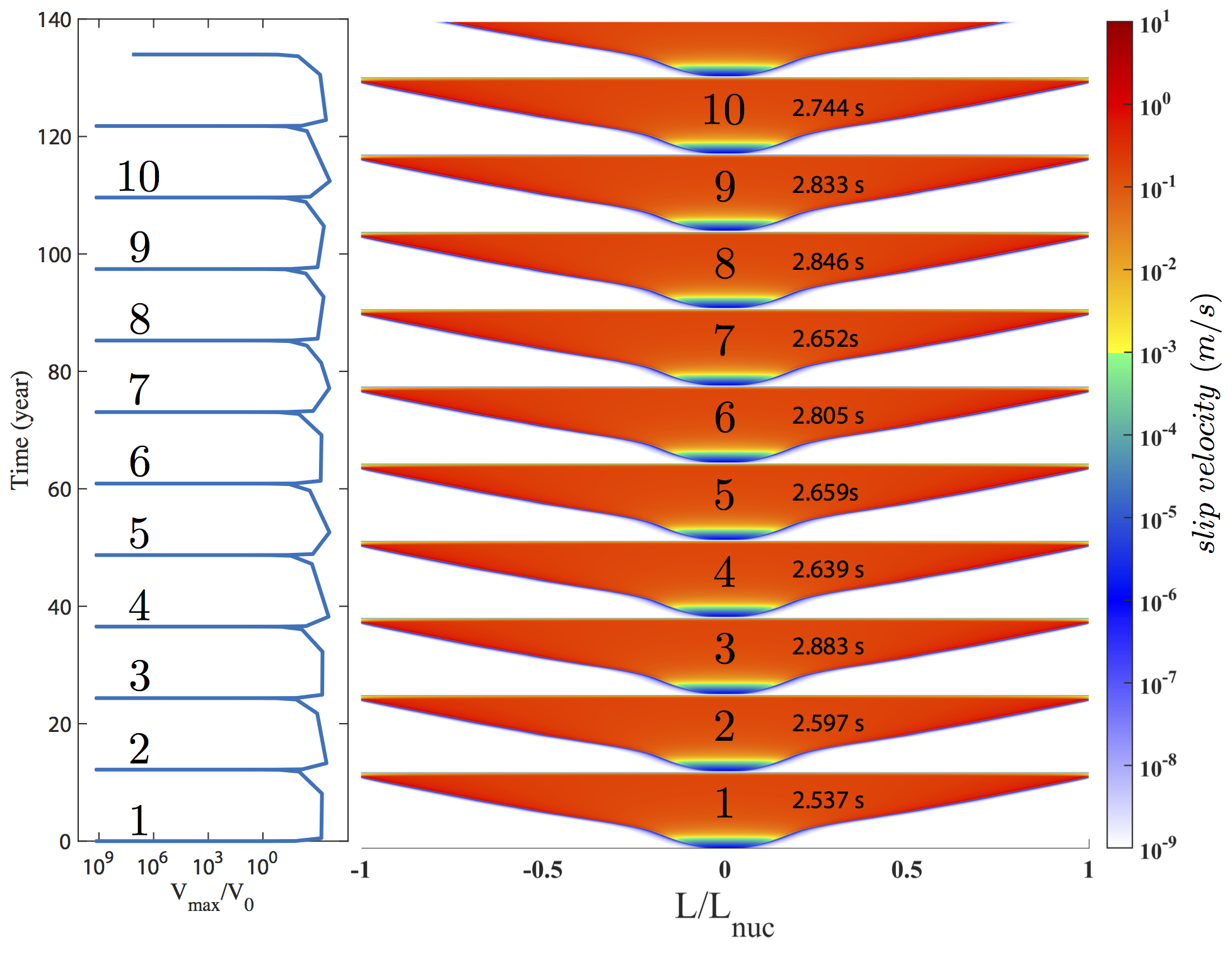}
\caption{Same length and friction parameter as in figure \ref{example}. Here $L/L_{nuc} = 2$ and $a/b = 0.9$. To avoid any artefact from initial conditions, the first 5 events of the simulation shown were removed.  Left panel shows the maximum slip velocity for fault 1 (blue). Right panel represents the space-time evolution of slip velocity on the faults. The highlighted duration of events corresponds to the time when the slip velocity exceeds 1mm/s for the first time to the time when it decelerates below 1mm/s.}
\label{SuppFigure2}
\end{figure}

\noindent\textbf{Single fault simulation}\\
The behavior of a single fault system with same length and friction parameters as in figure \ref{example}. As expected, the behaviour is periodic, without any complexity (Figure \ref{SuppFigure2}). Earthquakes nucleate at the same location (the center of the fault) and last for the more or less the same duration (close to 2.7s). 

\vspace{1cm}
\noindent\textbf{Estimating stress drops for Earthquakes and Slow-Slip Events}\\
Based on the work of H. Noda and colleagues \cite{noda2013b} we define three different measures of average stress-drops.

\noindent Seismological stress-drop is related to the seismic moment, $M_{0}$ and the length of the rupture, $L_{rup}$, as 
\begin{equation}
\Delta \sigma_{M} = \dfrac{4}{\pi} \dfrac{M_{0}}{L_{rup}^{2}}
\end{equation}

\noindent Spatial average of the stress-drop is evaluated using,
\begin{equation}
\Delta \sigma_{A} = \dfrac{\int_{L_{rup}}\Delta\sigma(l)~dl}{L_{rup}}
\end{equation}

\noindent And finally, the slip average stress drop is evaluated using the slip distribution, $\Delta u(x)$, as 
\begin{equation}
\Delta \sigma_{E} = \dfrac{\int_{L_{rup}}\Delta\sigma(l)\Delta u(l)~dl}{\int_{L_{rup}}\Delta u(l)~dl}
\end{equation}

\vspace{1cm}
\noindent\textbf{Length scales in the model:}\\
The nucleation length is defined in equation (\ref{nucleation}). The table below illustrates the actual length of the faults, $L$, (1--4 $L_{nuc}$) and the distance between them, $D$, (0.1--0.5 $L_{nuc}$) for various values of normal stress, $\sigma_n$, rate and state parameter, $a/b$ and critical slip distance, $D_{c}$.

\begin{table}[h!]
\centering
\begin{tabular}{|l| r| r| r| r| r|}
\hline
$\hspace{4.3cm} \bf a/b$& \bf 0.7 & \bf 0.8 & \bf 0.85 & \bf 0.9 & \bf 0.95\\ \hline 
$\sigma_n$ = 10MPa; $D_c$ = 0.1mm; $L$ (km) &0.2--0.6 & 0.5--1.4 & 0.8--2.5 & 1.9--5.7 & 7.6--22.9 \\
$\sigma_n$ = 10MPa; $D_c$ = 0.1mm; $D$ (m) &21--106 & 48--239 & 85--424 & 191--955 & 764--3820 \\ \hline
%%%%%%%%%%%%%%%%%%%%%%%%%%%
$\sigma_n$ = 50MPa; $D_c$ = 0.1mm; $L$ (km) &0.04--0.1 & 0.1--0.3 & 0.2--0.5 & 0.4--1.1 & 1.5--4.6 \\
$\sigma_n$ = 50MPa; $D_c$ = 0.1mm; $D$ (m) &4--21 & 10--48 & 17--85 & 38--191 & 153--764 \\ \hline\hline
%%%%%%%%%%%%%%%%%%%%%%%%%%%
$\sigma_n$ = 10MPa; $D_c$ = 1.0mm; $L$ (km) &2.1--6.4 & 4.8--14.3 & 8.5--25.5 & 19.1--57.3 & 76.4--229.2 \\
$\sigma_n$ = 10MPa; $D_c$ = 1.0mm; $D$ (m) &212--1061 & 477--2387 & 849--4244 & 1910--9549 & 7639--38197 \\ \hline
%%%%%%%%%%%%%%%%%%%%%%%%%%%%
$\sigma_n$ = 50MPa; $D_c$ = 1.0mm; $L$ (km) &0.4--1.3 & 1.0--2.9 & 1.7--5.1 & 3.8--11.5 & 15.3--45.8 \\
$\sigma_n$ = 50MPa; $D_c$ = 1.0mm; $D$ (m) &42--212 & 95--477 & 170--849 & 382--1910 & 1528--7639 \\ \hline
\end{tabular}
\end{table}
%%%%%%%%%%%%%%%%%%%%%%%%%%%%%%%%%%%%%%%%
%%%%  END OF DOCUMENT
%%%%%%%%%%%%%%%%%%%%%%%%%%%%%%%%%%%%%%%%
\end{document}